# Design and Simulation of Memristor-Based Artificial Neural Network for Bidirectional Adaptive Neural Interface

**Sergey Shchanikov[1*], Anton Zuev[1], Ilya Bordanov[1], Sergey Danilin[1], Dmitry Korolev[2], Alexey Belov[2], Yana Pigareva[3], Alexey Pimashkin[3], Alexey Mikhaylov[2], Victor Kazantsev[3]**

[1]Department of Information Technologies, Vladimir State University, Vladimir, Russia

[2]Research Institute of Physics and Technology, Lobachevsky University, Nizhny Novgorod, Russia

[3]Department of Neurotechnologies, Lobachevsky University, Nizhny Novgorod, Russia

* **Correspondence:**
Sergey Shchanikov
seach@inbox.ru

## Abstract

A key problem at hardware implementation of artificial neural networks based on memristors (ANNM) is to ensure the required accuracy of their operation at the transition from models to real fabricated memristive devices. Due to a number of factors, such as the imperfections in state-of-the-art memristors and memristive arrays, ANNM design and tuning methods, additional computation errors occur during the process of ANNM hardware implementation. This article proposes a general approach to the simulation and design of a multilayer perceptron (MLP) network on the basis of cross-bar arrays of metal-oxide memristive devices. The proposed approach uses the ANNM theory, tolerance theory, simulation methodology and experiment design. The tolerances analysis and synthesis process is performed for the ANNM hardware implementation on the basis of two arrays of memristive microdevices in the original 16×16 cross-bar topology being a component of bidirectional adaptive neural interface for automatic registration and stimulation of bioelectrical activity of a living neuronal culture used in robotics control system. The ANNM is trained for solving a nonlinear classification problem of stable information characteristics registered in the culture grown on a multi-electrode array. Memristive devices are fabricated on the basis of a newly engineered Au/Ta/ZrO$_2$(Y)/Ta$_2$O$_5$/TiN/Ti multilayer structure, which contains self-organized interface oxide layers, nanocrystals and is specially developed to obtain robust resistive switching with low variation of parameters. An array of memristive devices is mounted into a standard metal-ceramic package and can be easily integrated into the neurointerface circuit. Memristive devices demonstrate bipolar switching of anionic type between the high-resistance state and low-resistance state and can be programmed to set the intermediate resistive states with a desired accuracy. The ANNM tuning, testing and control are implemented by the FPGA-based control subsystem. All developed models and algorithms are implemented as Python-based software.

## 1    Introduction

Currently, many groups of researchers and manufacturers of computing equipment around the world are conducting large-scale R&D in the area of artificial cognitive systems, which are necessary to implement neuromorphic computing devices for neurorobotics and artificial intelligence, in solving topical problems of neuroprosthetics and neurorehabilitation (Schuman et al., 2017; Romano et al., 2019; Xia and Yang, 2019). This situation is caused by their potential advantages in accuracy, fault tolerance, performance, reliability and energy consumption over traditional information processing devices (Zidan et al., 2018) with von Neumann architecture.

However, the nominal quality of artificial neural networks (ANN) operation achieved at the stage of computer design is reduced in real operation conditions in many cases, sometimes followed by a complete loss of operability. The reason for this is the inevitable influence of internal and external physical and informational factors destabilizing the ANN operation, as well as manufacturing and operational errors of parameters of their implementation platform elements (Yeung et al., 2010; Torres-Huitzil and Girau, 2017).

The research results and reviews published by leading groups (Schuman et al., 2017; Zidan et al., 2018; Xia and Yang, 2019) show that the most prospective architecture of artificial cognitive systems for various applications is a neural network architecture with the state-of-the-art memristor-based hardware components in the form of simple thin-film structures that adaptively change their resistance depending on the application of voltage or current (Chua, 1971; Strukov et al., 2008). In recent years, significant progress has been made in fabrication of large arrays of memristors in cross-bar topology integrated with analog-digital circuits for hardware implementation of basic vector-matrix multiplication operations (Kataeva et al., 2019), as well as compact functional board-integrated MLP circuits (Bayat et al., 2018; Mikhaylov et al., 2018).

The active development of this direction and the analogy in the principles of constructing neural network architectures and living brain networks makes it possible to make the next step towards neurohybrid systems at the interface between artificial memristive systems and natural living systems in the form of cellular neuronal cultures or brain tissues (Chiolerio et al., 2017) to address the current challenges of robotics, artificial intelligence and medicine.

Within the framework of the leading European R&D programs in the period from 2013 to 2015, a number of projects have been launched aimed at creating neuromorphic chips based on memristors, as well as coordinating the efforts of European teams and industrial companies in this broad interdisciplinary field. Among these projects, it is necessary to highlight the RAMP project (Real neurons-nanoelectronics Architecture with Memristive Plasticity, http://www.rampproject.eu), the authors of which stated that, for the first time, natural and artificial neurons would be merged into a unique entity, namely a bio-hybrid adaptive interface based on memristors would be created for new computing systems and robotics. An important feature of the RAMP project was that the authors attributed to the most important tasks not only the creation of a biosimilar neurochip, but also the direct use of memristors in the registration and processing of bioelectric activity (Gupta et al., 2016, 2018). In the article (Serb et al., 2017) the project participants reported successful testing of a geographically distributed bio-hybrid neural network in which the spike-like signal of artificial neurons on a neuromorphic chip "passed" through memristors and "stimulated" the bioelectric activity of living neurons. The response of living neurons then returned via the Internet, "passing" through memristors, to the neuromorphic chip. The authors called their system "Internet of Neuroelectronics" (IoN). Arbitrary signals with different frequencies were applied to the memristor-based chip, and if it satisfied the plasticity conditions (a predetermined learning rule based on spike frequency modulation), the special setup began to generate programming pulses that set a predetermined resistance level on the memristor. This resistance value (memristor weight) was converted into parameters of the pre-developed protocol for stimulation of living neurons. That is, no physical connection of the elements was implemented, it was mediated and programmed. The only living elements in that network were the brain cells with a non-predictable response to stimulation. However, from the viewpoint of testing well-coordinated interaction of heterogeneous equipment via a specific protocol, that work was very significant and showed the relevance of distributed or compact neural interfaces based on memristive devices and systems.

It should be noted that this research direction is the most ambitious part of a more general research front devoted to the creation of neurointerfaces (Vassanelli and Mahmud, 2016). Currently, complex circuits in the form of neuromorphic processors and spiking neural networks or special mathematical models are used to implement real-time bi-directional neural interfaces (Hogri et al., 2015; Boi et al., 2016; Vassanelli and Mahmud, 2016; Buccelli et al., 2019). To the best of our knowledge, no such systems based on memristors have been implemented so far other than those planned in the mentioned RAMP project. The motivation of present work is related to the fact that it is the brain-like memristive system that will provide the highest degree of adaptability, energy efficiency and scalability that are required to implement compact and efficient neurointerfaces. In this work, it is for the first time proposed to develop a very simple perceptron scheme (well controlled and described theoretically) from the side of artificial electronic system, and a spatially separated culture with reproducible spatiotemporal patterns of response to stimuli from the side of living system.

The main research task in this direction is a search for circuit, design and technology solutions to the implementation of artificial neural networks based on memristors (ANNM), which will make it possible to bring their main parameters and characteristics closer to the potentially achieved values (Bayat et al., 2018; Mikhaylov et al., 2018; Emelyanov et al., 2019; Kataeva et al., 2019).

As follows from the level of ANN theory achieved to date (Galushkin, 2007) and the analysis of published scientific and technical papers carried out by the authors of this article, it is currently impossible to develop analytical methods of the ANNM design. This is due to the hardly formalizable, multidimensional, nonlinear and stochastic nature of ANNM and the problems to be solved, as well as certain destabilizing factors, physical and information processes in them.

The analysis of literature shows that the theory of design, fabrication and operation of ANNM is at an early stage of development (Lanza et al., 2019). An important section of this theory is the creation of methods and algorithms for automated engineering design of ANNM, as well as determining and ensuring the required values of the quality metrics of their operation regulated by national and international standards.

Key problems hindering the implementation of ANNM are:

- shortcomings in the ANNM theory;
- drawbacks of design and fabrication technologies of nanoscale electronic components and chips based on them;
- insufficient understanding of unique electrical and physical properties of materials and structures used to fabricate memristors (Chua, 2018).

In addition to the aforementioned reasons, the authors of the present paper have investigated and demonstrated in a number of research reports (Danilin et al., 2015, 2016) that the ANNM should be studied, designed, fabricated and operated as integrated physical and informational systems implemented with trainable hardware and software.

Many fault models of digital circuit elements used for hardware implementation of ANN and methods of their accounting are considered in the paper (Torres-Huitzil and Girau, 2017). They can be used in the design of digital parts of ANN, but at the same time, one of the most important unsolved problems in this area is the development of algorithms and approaches to determine and ensure the required operation accuracy of ANNM implemented on analog circuits elements, as well as determine and provide their fault tolerance (FT) and reliability (Danilin et al., 2015, 2016). At the informational level, FT is associated with the sensitivity of ANN, the analysis methods for which are given in the paper (Yeung et al., 2010). The sensitivity analysis is insufficient from the point of view of a system approach to the ANNM design, because the perturbations of their components at the informational level are indissolubly related to the errors of their components at the physical level.

A substantial contribution to the investigation of the influence of memristors non-idealities on the ANNM inference accuracy was made by the authors of paper (Mehonic et al., 2019). They showed that the main parameters of memristors (such as the high- and low-resistance device states, a finite number of discrete resistance states, device-to-device variability, etc.) affect the ANNM operation accuracy. Because of this, a quantitative criterion is needed that makes it possible to evaluate the FT for ANNM with different sets of weights or to calculate such values of weights that provide maximum FT.

This paper is devoted to the development of the pre-production design of memristor-based artificial neural network for bidirectional adaptive neural interface. The authors propose and apply a general approach, method and specific algorithm for the simulation and design of ANNM synapses with the required accuracy. These synapses are implemented in hardware with the integrated metal-oxide memristive nanostructures. A new version of the quantitative criterion of the ANNMs operation accuracy is proposed taking into account the tolerances for their parameters.

## 2    Materials and methods

A project of ANNM is a set of structural and functional models, necessary and sufficient to solve the tasks assigned in terms of reference. During the engineering design of ANNM in compliance with international standards in the field of electronics design (IEC, ISO), the accuracy, fault tolerance, reliability and performance of operation in normal conditions and under the influence of destabilizing factors should be determined.

### 2.1    General approach to ANNM design

The general approach to engineering design, development, study and operation of ANNM (in particular, to the solution of the problem posed in the article) is based on the theory of system analysis, methodology of simulation and the design of experiment (Shannon, 1975), neural networks theory (Galushkin, 2007), and includes the following main terms and assumptions.

ANNMs should be studied, designed and fabricated as integrated physical and informational systems (see **Figure 1A**) implemented with the use of trainable hardware and software. At the informational level, the ANNMs perform information processing. At the physical level, information carriers are signal parameters (such as amplitude, frequency, duration, etc.) or their combinations. Informational and physical, internal and external factors destabilize the ANNM normal operation act jointly in most cases (Danilin et al., 2015).

The application of simulation means that all ANNM models should be divided into several structural and functional levels (see **Figure 1A**):

- At the system level, the ANNM should be considered from the standpoint of ability to perform assigned tasks as a "black box" with the values of quality characteristics given in the specification (accuracy, fault tolerance, reliability, performance, power consumption, etc.).
- The level of subsystems is represented by the models of information processing. The ANNM model at this level consists of a group of models such as mathematical, algorithmic, or software (for example, an artificial neuron, as a subsystem of ANNM, sums weighted inputs to produce and pass an output through transfer function).
- The level of devices is represented by models of signal processing (circuit diagrams). Algorithms for signal and information processing usually do not match, and their number is different.
- At the component level, the models of circuit elements are considered in part of physical processes that occur in them (different types of memristor models (ideal model, physics-based models, phenomenological models, stochastic and probabilistic models (Brivio et al., 2019; Pershin and Di Ventra, 2020).

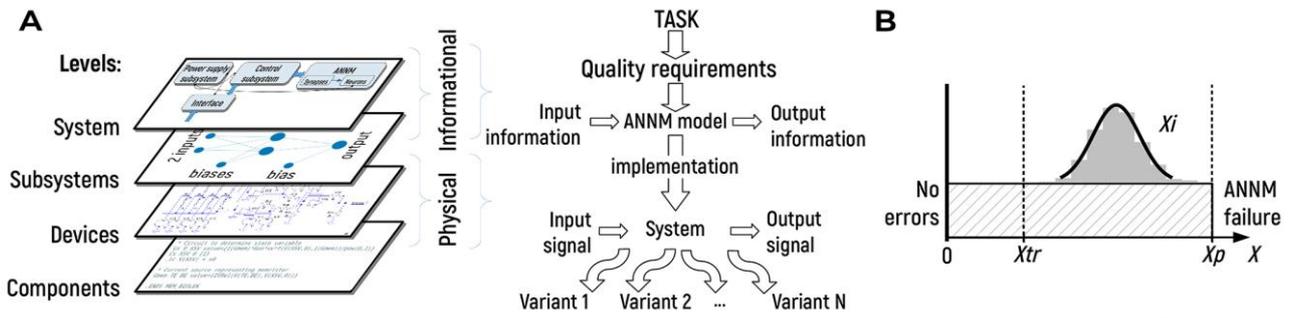

**Figure 1 |** The general approach to ANNM design. **(A)** The use of different types of ANNM models in their design process, which consists of the following steps. Before creating a system, a task is formulated. Then, one should define requirements for the task. There are always a lot of them, but some are more important. For example, for a pattern recognition system, low probability of error is more important than low power consumption. On the one hand, ANNMs perform information processing. Therefore, the operation accuracy is the main requirement for them. On the other hand, while choosing between different variants of model implementation, one can say that ANNM perform signal processing. **(B)** Tolerance ranges for ANNM. Distributions of values of all ANNM parameters should be inside valid boundaries.

At the basis of simulation models are algorithms that determine the logic of operation of ANNM and their components at each level of the hierarchy, as well as assessment and optimization methods for the values of quality characteristics.

The technology for implementing this approach includes the following steps:

- Choosing the level of hierarchy at which it is necessary to carry out the simulation.
- Setting the range and step of changing the physical and (or) informational parameters of destabilizing factors.
- Design of experimental plans and simulation.
- Approximation and visualization of simulation results.
- Optimization of parameters and characteristics of components of the developed ANNM in order to provide the necessary quality.

At the physical level, a signal conversion is performed by electronic components which inevitably have variations in their electrical parameters in comparison with nominal values. These variations are caused by various reasons: physical, circuit-related, design-related, process-related, etc. That is why after the ANNMs hardware implementation, their operation accuracy decreases relative to the theoretical value. The considered set of models makes it possible to carry out the engineering design activities, implementation, and study of ANNM.

## 2.2 Method for defining ANNM tolerances

All possible destabilizing factors causing the degradation of accuracy of ANNM operation can be conveniently classified into internal (device-to-device variability, stuck-at devices, noise and fluctuations of electrical parameters, etc.) and external (noise and errors in the input signals and information, interference and errors of power sources, etc.). For providing the required accuracy of ANNM operation, it is necessary to take into account the influence of destabilizing factors specific for a hardware implementation under consideration. As a result of this work, the maximum permissible levels of these destabilizing factors can be determined. Methods and tools of providing active and passive fault tolerance of ANNM to destabilizing factors are considered in more detail in (Yeung et al., 2010; Torres-Huitzil and Girau, 2017; Mehonic et al., 2019).

In compliance with international standards, the permissible limits of variations in parameters and characteristics of electronic components and devices must be specified with certain tolerance ranges (tolerances) during development, manufacturing production and operation. Unreasonably rigid tolerances cause an increase in cost, complicate a technology and prolong the term of product development. With regard to ANNM, by defining optimal tolerances for tasks with various accuracy requirements one can reasonably choose different in their complexity and cost: types and architectures of ANNM, industrial memristor fabrication processes, circuits, design solutions for combining them into arrays (cross-point, cross-bar, discrete memristive devices), tuning (learning) algorithms, etc. From all points of view, it is more appropriate to define tolerances at the design stage.

Talking about ANNMs, the values of their performance metrics can be divided into three types: achieved while training a model ($X_{tr}$), permissible for solving the practical task ($X_p$) and obtained after hardware implementation ($X_i$). Here $X$ is one of the metrics used in machine learning (for example, Mean Squared Error or some other). A tolerance range for ANNM output parameters is shown in **Figure 1B**.

Then, one should divide the determination of tolerances into two tasks – the analysis and the synthesis of them:

- the analysis of tolerances makes it possible to investigate whether the neural network operates with the required accuracy in the presence of the known limits of errors of its informational or physical parameters;
- the synthesis of tolerances makes it possible to determine the limits of errors of informational or physical parameters allowing the neural network to operate with the required accuracy.

At the known probability density function of variations in the ANNM parameters (see **Figure 1B**), it is appropriate to use probabilistic methods to define tolerances. This may be done with the use of simulation and the design of experiments. According to (Danilin et al., 2015, 2016), to define the tolerances for ANNM informational parameters, one should:

- Choose a performance metric for the developed ANNM (Sum of squared error (SSE), Mean of squared error (MSE), Mean absolute error (MAE), etc.) and set its permissible value $X_p$ for the task to be solved.
- Create the model of ANNM with the given properties (architecture, structure, learning algorithm, etc.) at the informational level.
- Train the ANNM model with the selected algorithm using software ("ex-situ") until $X_{tr}$ becomes less than $X_p$ or set the parameters obtained after hardware training (tuning) ("in-situ") to the simulation model.
- Perform the simulation of variations in the ANNM parameters caused by hardware imprecision and external or internal noise.
- Store the current value of the ANNM operation accuracy $X_{i,j,k,l,f}$ for each simulation iteration.
- Assess whether the ANNM is operating within the tolerance – operation accuracy obtained after hardware $X_{i,j,k,l,f}$ implementation should be better than the permissible $X_p$ value.

As a result of these steps, a tolerance analysis should be performed. For a tolerance synthesis, the steps 4 - 7 should be repeated several times according to the experimental plan. In each experiment, one should define such values of limits, at which ANNM operates with the permissible accuracy. These values are the tolerance ranges for the parameters of ANNM components under investigation.

The tolerances for physical parameters of the ANNM components should be in accordance with the tolerances of informational parameters.

## 2.3 Neural interface description

The tolerances analysis and synthesis process was performed for the ANNM implemented with arrays of memristive microdevices in the original 16×16 cross-bar topology (Gryaznov et al., 2018) being a component of bidirectional adaptive neural interface for automatic registration and stimulation of bioelectrical activity of a living neuronal culture used in robotics control system. It can be used to create compact neuromorphic controllers for neurorobotics on the basis of integration of ANNM into the interface circuitry of a cultural neural network of dissociated brain cells (see **Figure 2A**). The ANNM is trained for solving a nonlinear classification problem of stable information characteristics registered in the culture cultivated on a multi-electrode array (MEA). However, the ANNM functionality is not only limited to the neuronal activity decoding, but includes the active control of their stimulation. In particular, the proportional characteristics resulting from the ANNM classification are used for gradual manipulation (stimulation with various amplitudes and numbers of electrodes) stabilizing the desired culture functional activity (see **Figure 2B**).

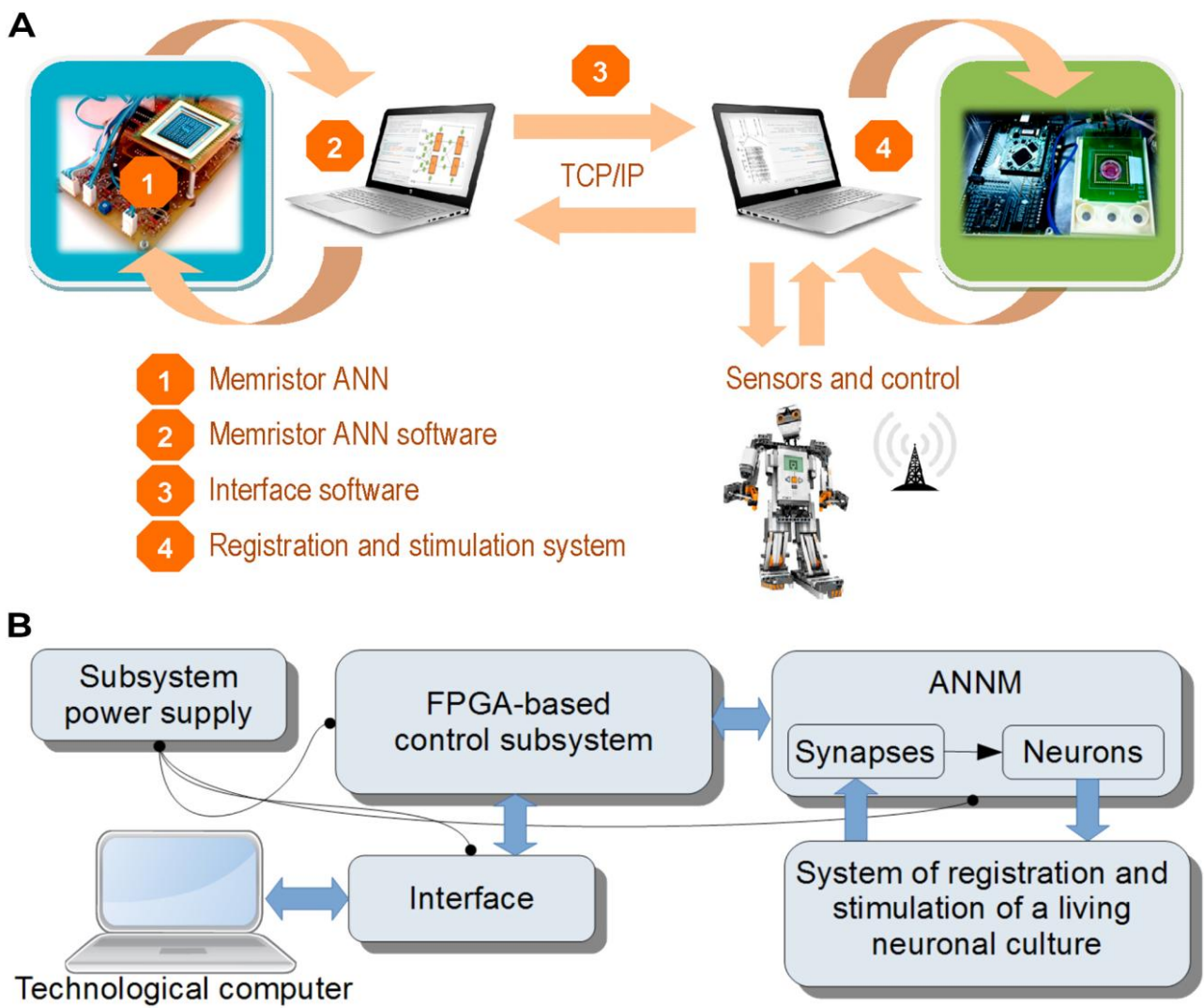

**Figure 2 |** Bidirectional adaptive neural interface. **(A)** Functional diagram. **(B)** Place of the ANNM in the bidirectional adaptive neural interface. The ANNM task – classification of the patterns of living culture signals. The ANNM tuning, testing and control are implemented with the FPGA-based control subsystem.

### 2.3.1 Microfluidic device fabrication

To obtain the spatially ordered neuronal cultures during their growth on MEA, the microfluidic chips were fabricated via polydimethylsiloxane (PDMS) moulding techniques. Standard two-layer lithography was used for mould fabrication. The microfluidic chip consists of two chambers for cell cultivation and 8 microchannels providing unidirectional axon growth from Source chamber to Target chamber. Microchannels design was based on a sequence of 3 or 4 triangle segments that facilitated the directed axon growth (see **Figure 3A**).

The surfaces of the prepared PDMS chips were mounted with MEAs, which were coated with the adhesion-promoting polyethyleneimine molecules at the concentration of 1 mg/mL (Sigma-Aldrich, USA) and laminin at the concentration of 20 mg/mL (Sigma-Aldrich, USA). The chips were manually aligned with the MEA, which was composed of 60 electrodes (TiN electrodes, diameter 30 µm with 200 µm in between, Multichannel Systems, Germany), via a three-dimensional mechanical micromanipulator under a binocular. Furthermore, 14 or 24 electrodes were placed in each chamber, 24 or 32 electrodes were placed in the microchannels (3 or 4 electrodes in each of 8 microchannels) (see **Figure 3B**). We used reversible bonding for MEAs to prevent damage of the electrodes. After the PDMS chips were mounted to the MEA, it was cured in an oven at 80 °C for 30 min.

### 2.3.2 Cell culturing

Hippocampal cells were dissociated from embryonic mice (E18) and plated in the cell chambers of PDMS chips at an initial density of approximately 7,000–9,000 cells/mm$^2$. Mice were euthanized via cervical dislocation according to protocols approved by the National Ministry of Health for the care and use of laboratory animals. The protocol was approved by the Committee on the Ethics of Animal Experiments of the Nizhny Novgorod State Medical Academy. All efforts were made to minimize suffering. For culturing procedure details see (Pimashkin et al., 2013). The cells were cultured under constant conditions of 37 °C, 5% $CO_2$ in a humidified cell culture incubator (MCO-18AIC, SANYO, Japan).

### 2.3.3 Electrophysiology

Extracellular potential measurements were performed after 20 day *in vitro* when two cultures in the microfluidic device were already coupled by the axons through the microchannels and generated spontaneous activity. Detection of the recorded spikes was based on the threshold calculation of the signal median as described in our previous studies (Pimashkin et al., 2011, 2013). Signal analysis and statistics were performed with custom made software in MATLAB.

Stimulation through the MEA was performed with the STG-4004 pulse generator (Multichannel Systems, Germany). Series of 30 stimuli were applied to each of four randomly chosen electrodes in the Source chamber. Low-frequency stimulation consisted of biphasic voltage pulses ±800 mV, 260 µs per phase, positive first, intervals between stimuli were 3 s. Signals from random 4 electrodes placed in the middle of microchannels were recorded by the MEA system (Multichannel Systems, Germany) at a sample rate of 20 kHz. An example of stimulus-evoked activity recorded on one of the electrodes is shown in **Figure 3C** and the typical histogram of spike timings in response to 30 stimuli is shown in **Figure 3D**. Network responses to stimulation of different electrodes are shown in **Figure 3E**.

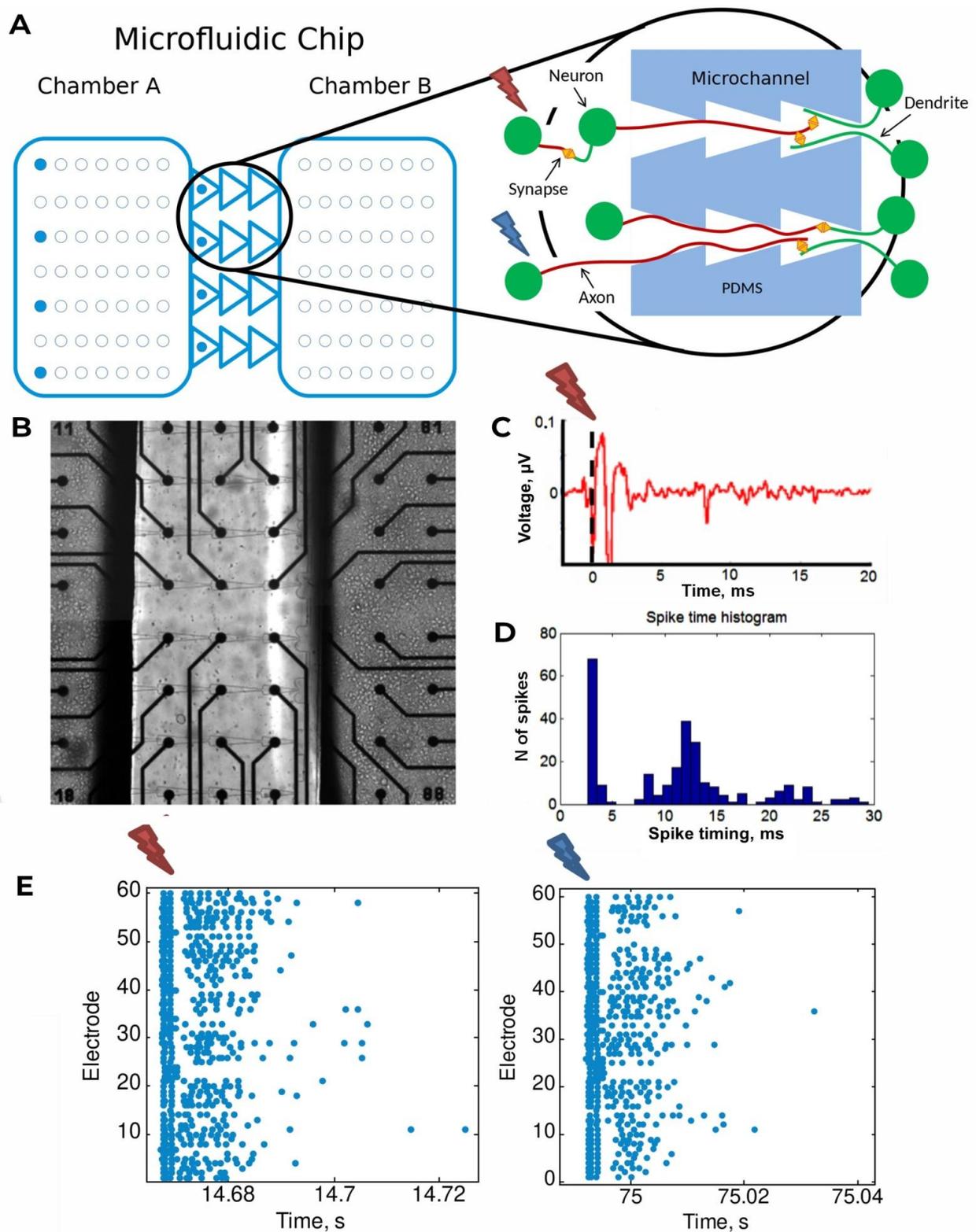

**Figure 3 |** Microfluidic chip. **(A)** Schematic illustration of neural network grown in microfluidic chip with ordered connectivity between two subnetworks. **(B)** Neuronal cells plated in two chambers of microfluidic chip and connected through microchannel for unidirectional axon growth (DIV 15) and combined with MEA, scale bar = 200 um. **(C)** Example of response recorded on electrode in response to applied stimulus: direct axonal spike (<5 ms) and synaptically evoked response (>5ms). **(D)** Histogram of spike timings in response to 30 stimuli. First sharp peak is associated with direct axonal spike, second – with synaptically evoked response with ~3 ms jitter. **(E)** Examples of network response to stimulation of two different sites. Every dot corresponds to the spike recorded on the electrode.

## 2.4 Memristive devices description

Memristive devices were fabricated on the basis of a newly engineered Au/Ta/ZrO$_2$(Y)/Ta$_2$O$_5$/TiN/Ti multilayer structure, which contains self-organized interface oxide layers, nanocrystals and is specially developed to obtain robust resistive switching with low variation of parameters (Tihov et al., 2020). An array of memristive devices is mounted into a standard metal-ceramic package (see **Figure 4A**) and can be easily integrated into the neurointerface circuit. Memristive devices demonstrate bipolar switching of anionic type between the high resistance state (HRS) and low resistance state (LRS). Both states are characterized by nonlinear current-voltage characteristics (CVC) (see **Figure 4B**) and low resistance variation obtained from CVC at the voltage of 0.5 V (see **Figure 4C**). It is worth noting that such nonlinear characteristics are appropriate for the formation of passive cross-bar arrays with a highest density per chip achieved at the moment (Bayat et al., 2018).

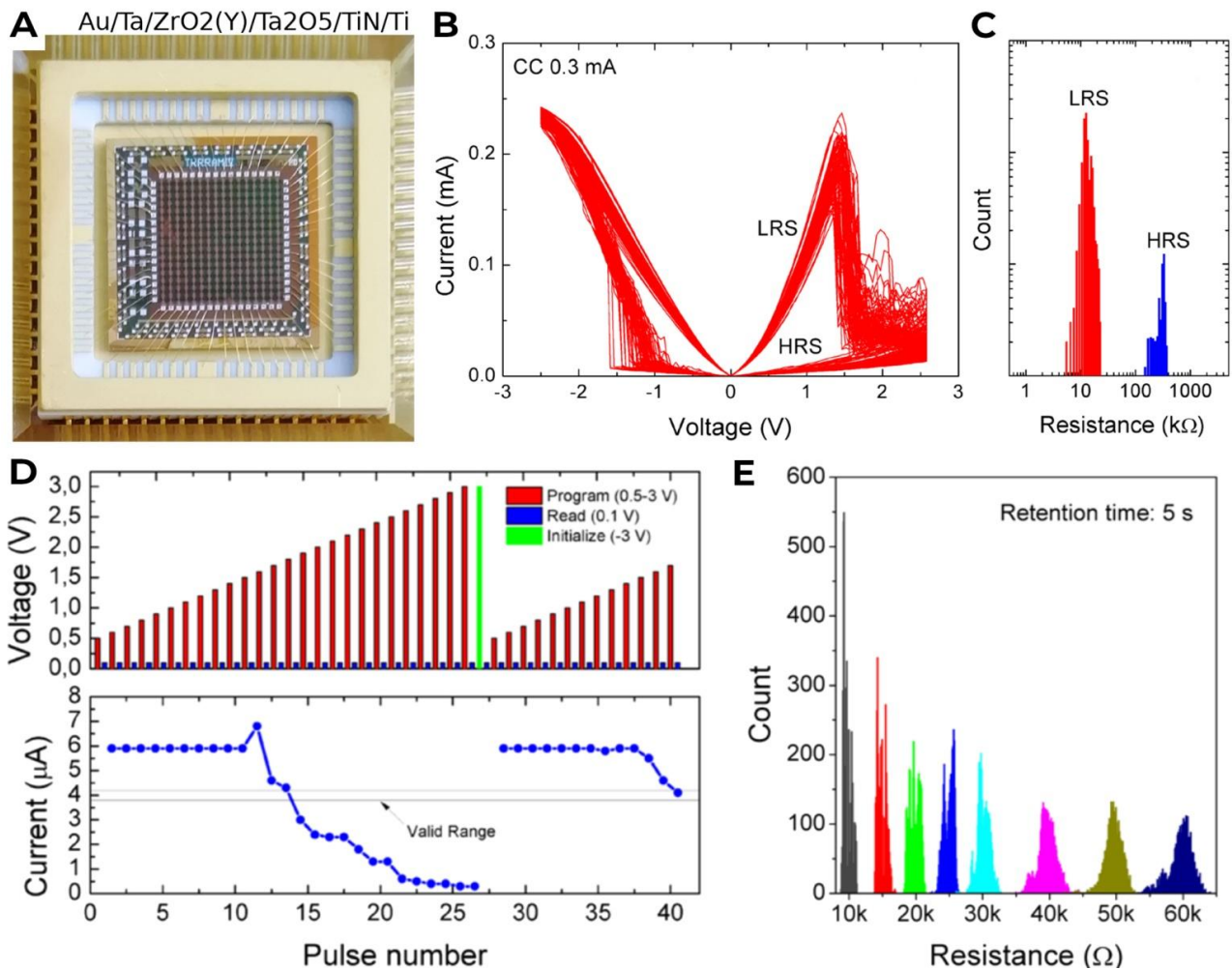

**Figure 4 |** Memristive microdevices. **(A)** Photograph of a packaged cross-bar array of memristive microdevices, **(B)** typical CVC in the bipolar switching mode (100 cycles) and **(C)** distributions of resistive states over 1,000 CVC cycles. **(D)** Voltage ramps applied to memristor and reading current response during programming by the developed active feedback algorithm. **(E)** Distributions of 8 different programmed resistive states.

To program the specified resistive states of memristive devices, which determine the corresponding synaptic weights in the ANNM, we used the voltage ramp pulse technique (Emelyanov et al., 2019) adapted to the peculiarities and parameters of resistive switching for the used memristive structure. A feature of the developed technique is the use of active feedback when setting a specified state with the required accuracy, which reduces the total number of actions on memristive device in comparison with other techniques. To set the programming modes, a range of possible states (resistance values) of the memristive device was determined from the analysis of its DC CVCs. Programming was carried out by applying voltage pulses with increasing amplitude in the range from 0.5 to 3 V (pulse duration of 1 ms) with reading the state after each pulse by applying a pulse with a small amplitude (0.1 V, 1 ms). In the event that the required resistance is not achieved with a given accuracy with an increase in the amplitude, the memristive device is rewritten to its initial state by applying an initializing voltage pulse (-3 V, 1 ms) with a current limit of 300 μA, and the above voltage pulse supply algorithm is repeated until the specified state is achieved. An example of a time series with two iterations of sequences of programming pulses and readable current values outside and inside a valid range is shown in **Figure 4D**.

**Figure 4E** shows a typical implementation of the proposed algorithm for programming 8 different resistive states in the range of 10-60 kΩ. To obtain the required statistics for building histograms, at the end of the programming procedure, an additional retention test was carried out by measuring the current through the memristor for 5 s with a constant voltage of 0.1 V. The programming procedure was repeated 20 times for each given resistance value. Despite the observed variation in the resistive states associated with the non-ideal retention characteristics, it can be seen from **Figure 4E** that the developed technique provides unambiguous programming of multiple states without overlapping their distributions.

The results of experimental measurements of the conductivity of memristors in the process of programming and their functioning in real operating conditions are necessary to set the parameters of the simulation model. Memristors were programmed with a given tolerance of ± 15% for any of the intermediate states. As can be seen from **Figure 4E**, after programming and reading, the conductivity error does not exceed the specified value. According to the Pearson's criterion, it was found that the distribution of errors obeys the normal law with an acceptable level of significance. These parameters are used later in the simulation.

## 2.5    ANNM description

### 2.5.1  Circuit design

A neuron is the basic element of ANN. It consists of synapses, a summator and an activation function. Memristors are used for the implementation of synapses in creating the ANNM, namely for the storage of weight values and the multiplication operation. The advantage of using memristors lies in the fact that any weight value can be set in a given range, since a memristor resistance is continuously variable.

There are several types of memristor physical models at the moment that can be used for the ANNM implementation (Pershin and Di Ventra, 2020). Single or paired memristors and memristor arrays can be used (Demin et al., 2015; Emelyanov et al., 2016). Single memristors are suitable for the testing purposes, but it is very difficult to develop even a medium-sized network (hundreds of memristors) using them. Difficulties arise from the problems with an individual synapses connection, interference in long transmission lines and other factors. For this reason, it is better to use the memristor arrays implemented on a chip to build the ANNMs that solve practical tasks.

Currently, existing arrays can be divided into two types: active (Yao et al., 2017) and passive (Park et al., 2015). A passive array is the set of memristors, wherein each memristor is connected to two common electrodes (column and row). The difference of active arrays is that the connection of each memristor to one of electrodes is performed through the active element, typically being a field effect transistor. It is worth noting that there is a problem of sneak currents for all memristor arrays (Hamdioui et al., 2014). Despite the fact that there is a solution in the form of memristor selective activation to resolve this problem in active arrays, such devices are more technologically complicated and their density on a chip will be theoretically much less due to the presence of active elements. Bias voltages applied to the column and row electrodes of a passive array according to a specific algorithm can be used to eliminate the influence of sneak currents on the states of memristors during a programming process (Park et al., 2015). In this work, passive crossbar arrays are used to implement the ANNM.

Multiplication of input data by weight is performed by converting the voltage of input signal into current, the memristor in this case acts as a current source. The current flowing through the memristor depends linearly on the input voltage according to the Ohm's law:

$$I_{IN} = \frac{U_{IN}}{R_M}, \tag{1}$$

where $U_{IN}$ is the voltage amplitude of input signal, $R_M$ is a memristor resistance. Thereby, the conversion factor is equal to $1/R_M$.

Consider the diagram of a single neuron (**Figure 5**). All memristors of separate row are connected to one inverting amplifier based on operational amplifier ($U_1$, $U_2$) summing currents at inverting input according to the Kirchhoff's point rule. The sum of all memristor currents is defined by the formula:

$$I_{SUM} = \sum I_k, \tag{2}$$

where $I_k$ is the current of $k$-th memristor in row.

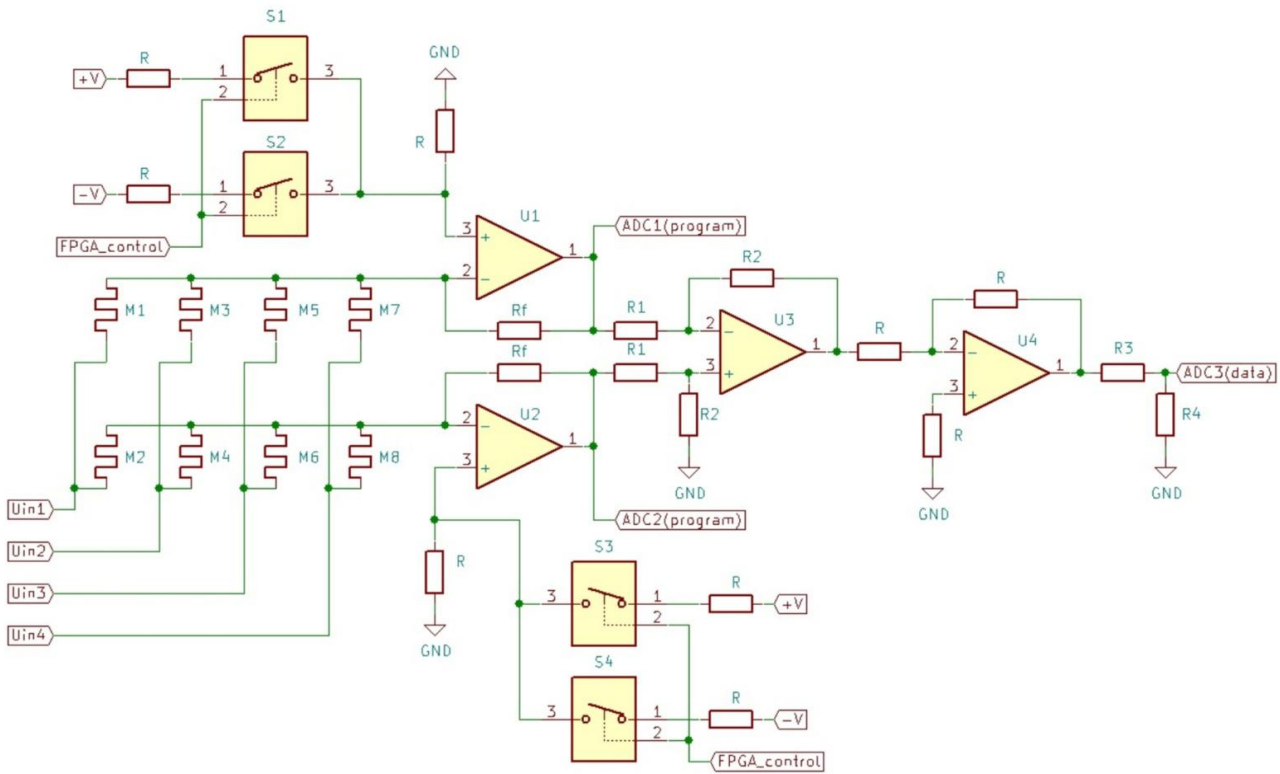

**Figure 5 |** Hardware implementation of the ANNM. A simplified circuit diagram of a neuron. Memristors-based synapses store weights and multiply input information by them. Input information at the physical level is voltage amplitude of an input signal. Multiplication is performed by Ohm's law. Addition is performed by Kirchhoff's rule. All synapses located in two adjacent lines belong to one neuron in order to obtain the bipolar values of weights. Neurons consist of differential amplifiers performed transfer functions.

Such solution is convenient, because negative feedback makes it possible to obtain not only a virtual ground on the inverting input, to which memristors are connected, but also different bias voltages by applying the appropriate voltage to the non-inverting input. Total current of all memristors and, accordingly, the current of each individual element are converted to the amplifier output voltage with a coefficient that is equal to the value of feedback resistor with a negative sign):

$$U_{SYN} = -I_{IN} \cdot R_F,\tag{3}$$

where $R_F$ is a value of feedback resistor.

Thus, the final conversion factor of an input signal will be defined as:

$$K = \frac{U_{SYN}}{U_{IN}} = -\frac{I_{IN} \cdot R_F}{R_M \cdot I_{IN}} = -\frac{R_F}{R_M}.\tag{4}$$

Only unipolar non-zero values of weights can be obtained with a single memristor that is a serious constraint in the implementation of a computer model. Therefore, in this paper, each synapse is implemented using a pair of memristors located in adjacent rows and related to the same column (data input). Thus, all memristors in a given pair of rows are associated with the synapses of one neuron. The output voltages of inverting amplifiers are differentiated to obtain bipolar values of the coefficients. A standard differential amplifier is used for this purpose ($U_3$). The weight value is defined as the difference between the voltage conversion coefficients coming to the inputs of this amplifier:

$$W = K_2 - K_1 = \frac{R_F}{R_{M1}} - \frac{R_F}{R_{M2}} = \frac{R_F(R_{M2} - R_{M1})}{R_{M1} \cdot R_{M2}},\tag{5}$$

where $R_{M1}$ is the resistance of a memristor related to the row whose inverting amplifier is connected to the inverting input of the differential amplifier, $R_{M2}$ is the resistance of a memristor related to the row whose inverting amplifier is connected to the non-inverting input of the differential amplifier.

The gain, moreover, a scaling factor of synapse weights, which can be controlled by changing the values of the resistors, is defined as:

$$K_{DIFF} = \frac{R_2}{R_1}.\tag{6}$$

A saturating linear transfer function is used as an activation function. It is easy enough to implement and can replace a sigmoid function, which is used in computer simulation, with little loss of accuracy. In the circuit, the activation function is implemented with an inverting amplifier ($U_4$), the gain of which characterizes a slope of linear section of the function. The saturation voltages of the amplifier characterize the saturation levels of the function.

Output signals need to be scaled before sending them to the next layer of the network. In the absence of strict performance requirements, this can be done with a common resistive voltage divider calculated in such a way that a scaling factor would be equal to the ratio of the maximum output voltage of the current network layer to the maximum input voltage of the next layer:

$$K_{SCALE} = \frac{U_{OUT.MAX}}{U_{IN.MAX}} = \frac{R_4}{R_3 + R_4}\tag{7}$$

It should be noted that the slope of an activation function will change in proportion to the scaling factor. The need to limit the voltage of input signals is due to the fact that memristors have a certain threshold voltage, below which the signal amplitude (regardless of the sign) does not change the state of memristors. For the arrays used in this work this level is ± 1.5 V. The voltage range for the signals representing data is limited to ± 1 V.

Programming of weight coefficients is carried out by the switching of memristors to the states with the specified values of resistances. For this purpose, a programming pulse with an amplitude above 1.5 V is applied to memristor. The higher the pulse amplitude, the higher is the resistance of the element. The resistance value depends on the state, in which the memristor was before the application of pulse, so the memristor state programming algorithm can be divided into several stages:

At first, the memristor is transferred to LRS by the applying a SET pulse with an amplitude of −3 V. For the SET mode, the current limit is set to 300 μA to avoid the situation when the memristor can go into a state with extremely low resistance. Too low resistance will cause the increase in current flowing through memristor above a critical point leading to the device damage or failure, as well as to the malfunction of DAC, from which programming pulses and data are applied.

At this stage, special attention should be paid to the problem of sneak currents. In the array, when the SET pulse is applied, the current is flowing not only through the selected memristor, but also through the chain of memristors sequentially connected in parallel with a given memristor. In this case, the voltage drop on individual memristors can be more than 1.5 V that will change their state. To prevent this situation, a bias voltage of −3 V is applied to all other rows and a bias voltage of − 1.5 V is applied to all other columns. This ensures that a voltage drop on any other than a selected memristor will be either 0 or 1.5 V and their state will not change. A bias voltage of −3 V is applied to all rows except the target one only when a SET pulse with an amplitude of −3 V is applied to the target column (i.e. it is not connected to ground), so the voltage drop across all memristors in this column is 0 V. In this case, the V/2 scheme is not suitable for the reason that during the memristor switching to LRS using SET pulse, current limitation is used, and an offset of −3 V ensures that no current flows through other memristors in the target column, which otherwise would cause early current limitation.

Then, RESET programming pulses of voltage with an amplitude of more than 1.5V with a specified step of change begin to be applied. After each programming pulse, the memristor resistance is read by applying a test pulse of 0.5 V not affecting its state through recording the output voltage of a summing amplifier by using the ADC and calculating the resistance by the formula:

$$R_M = \frac{U_{TEST} \cdot R_F}{U_{OUT}}, \tag{8}$$

where $U_{TEST}$ is the amplitude of a test pulse, $R_F$ is the value of a summing amplifier feedback resistor, $U_{OUT}$ is the output voltage of a summing amplifier.

If the memristor resistance corresponds to a target value within tolerance, the transition to the next element is carried out. If, after the test, the current value exceeds the required one, the memristor is again transferred to LRS and the procedure is repeated.

At this stage, the remaining rows and columns are supplied with a bias voltage of 1.5 V to solve the sneak current problem. In order to guarantee the maximum voltage drop on any other than the selected memristor is not higher than 1.5 V, a limit on the maximum amplitude of the programming pulses of 3 V is introduced. Thus, the voltage range of the programming pulses is [1.5; 3] V. During the application of test pulse, the remaining columns and rows are either connected to ground or zero bias is applied to them to avoid the presence of other memristors currents at the inputs of summing amplifiers.

At this technological level of memristor array fabrication, some elements may "freeze" in a certain state, which will not change after the programming pulses. In the work (Bayat et al., 2018), it is offered in this case to check all elements for the appropriate state after the fabrication of a chip. If such an element is found, its parameters are stored and then used in a computer simulation to calculate the bias value. In the present work, the arrays were checked after fabrication, too. If only one element of a pair of memristors implementing one synapse freezes, the coefficient is adjusted by programming the resistance of the second memristor relative to the first one. If both elements hang, then the ratio of their resistances is calculated and further counted as a fixed bias.

DACs with bipolar output voltage ranges are used to generate programming pulses and data signals of different amplitude and sign. The accuracy of the input data and the amplitude step of programming voltage ramp depend on the bit depth of the DAC. The data signal is formed digitally in FPGA and then fed to the DAC channels.

ADCs with bipolar input are used for reading values from output neurons in data processing mode, as well as for reading the output voltage from each summing amplifier in programming mode. Data signal from the ADC is transferred to the FPGA, where it is processed. There is no need for ADC at the output of neurons of the first layer because the training is performed in a simulation and readout of intermediate values is not required.

FPGA-based control subsystem in addition to communication with the DACs and ADCs performs signal switching. The bias voltages are formed on the rows by applying a reference voltage to the non-inverting inputs of summing amplifiers. Electronic switches controlled by FPGA are used to supply various biases. The advantage of using FPGA is that in contrast to microcontroller it is possible to build a parallel system in which the signals applied and values read at all outputs are carried out simultaneously. The simultaneous application of programming pulses and bias voltages is particularly important.

It is worth noting some requirements to operational amplifiers used for the ANNM implementation. Since the bias voltage of the memristor is set on the summing amplifier, it is necessary to provide a low input offset voltage and a low input bias current of the amplifier itself. For the same reason, it is desirable to use amplifiers with a low equivalent input noise voltage. At high values of these parameters, it would be necessary to reduce the values of bias voltages and therefore the voltage range of programming pulses.

In addition, the quality of the power subsystem will affect the accuracy of the entire network. For example, the system accuracy of an entire system will reduce at a significant level of ripple comparable to the step of a voltage change at the output of a DAC. Accuracy will also depend on the stability of the bus voltage supply, because the output values of the DACs, the saturation voltage of the amplifiers, the reference voltage and other parameters are changing with the changing of a supply voltage.

### 2.5.2 Training and test set

The main problem being solved by the ANNM under development is the recognition of patterns of neural activity that occur in response to stimuli applied to a neuronal culture in chamber A (see **Figure 6A**). The pattern of neural activity is spatio-temporal, as it is formed on the basis of information about the time of arrival of spikes in the channels of the microfluidic chip separated in space. In this study, the arrival times of the first four spikes in the four channels of the chip were recorded in response to four stimulation at sites $S_1$, ..., $S_4$ in chamber A during 50 ms. The experimental results showed that for these four cases, the deviation in spikes timings does not exceed $\pm 30\%$. Then, the timing data were approximated by a Gaussian function and the signal patterns were synthesized pseudo-randomly for four sources of stimuli (4000 signal patterns, 1000 for each stimulus) (see **Figure 6B**). This is necessary to increase the size of training and test sets for the developed ANNM.

**A**

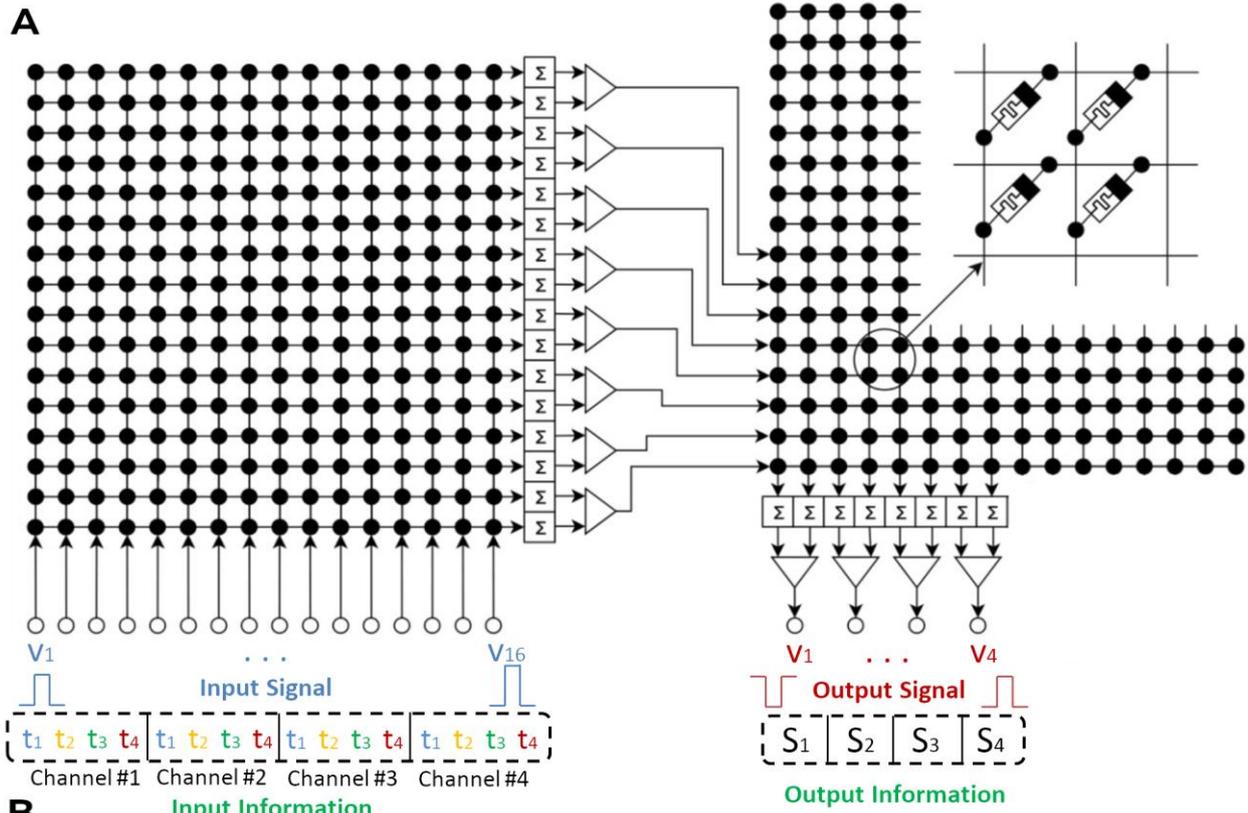

**B** Input Information

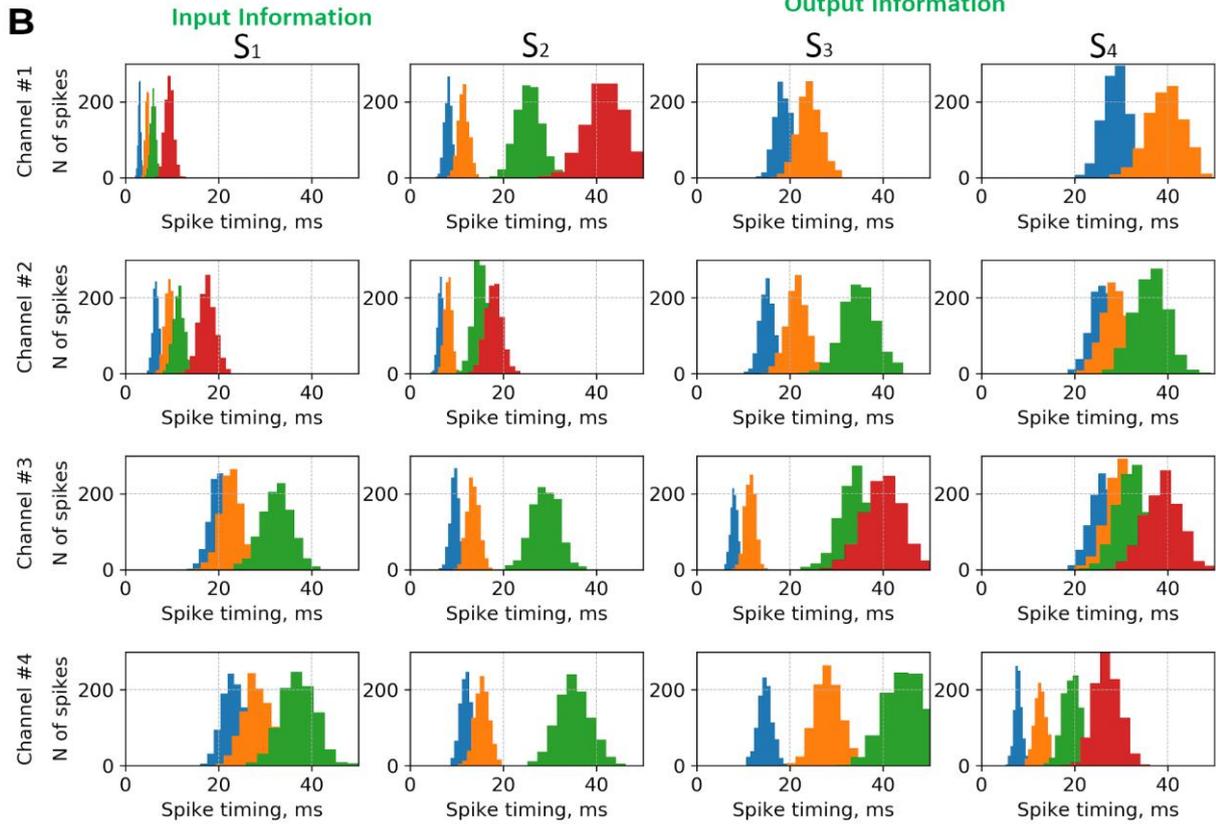

**Figure 6** | Training and test sets. **(A)** The block diagram of the ANNM under development. Colors of labels at the inputs of the MLP correspond to the colors of spikes timings ($t_1$, …, $t_4$) in histograms. **(B)** Spikes registered in four channels (Channel #1, …, Channel #4) after stimuli from four electrodes ($S_1$, …, $S_4$) in chamber A during 50 milliseconds. Histograms of spikes timings ($t_1$, …, $t_4$) which were generated on the basis of experimental measurements in the microfluidic chip. In some channels, the fourth spike came after 50 ms, so the training sample got a value of 50 for them (which corresponds to "1" after normalization).

In addition to the arrival times of spikes after stimuli from the sites $S_1$, ..., $S_4$, the patterns of neural activity that occurred spontaneously were registered and classified as extraneous ($S_r$). On the basis of extraneous patterns, 4000 pseudorandom sequences were synthesized for training and test sets. The total number of 8000 patterns was mixed pseudorandomly and divided into training set (735 for $S_1$, 760 for $S_2$, 724 for $S_3$, 754 for $S_4$, 3027 for $S_r$, total: 6000) and test set (265 for $S_1$, 240 for $S_2$, 276 for $S_3$, 246 for $S_4$, 973 for $S_r$, total: 2000).

Each signal pattern is a vector of 16 arrival times of spikes in four channels of a microfluidic chip during 50 ms. For normalizing the data and bringing them to a given range of voltage amplitudes, each element of training and test sets is divided by 50 and rounded to the specified bit depth of DACs. Bipolar 12-bit DACs with an output voltage range of ± 5V were used. Thus, all values of the training and test sets are rounded off with a step of 0.0025 V. Each stimulus site in chamber A is associated with a vector of four elements containing "1" at a position of current stimulus site S and "-1" at the remaining positions through the signal patterns on inputs. For spontaneous signals, a target vector consists of four "-1".

### 2.5.3 ANNM model

The ANN model has MLP architecture (see **Figure 7A**). Due to the fact that the signals registered in microchannels are not orthogonal and can have high cross-correlation, the task is not linearly separable, so there is a need to use a double-layer architecture. At the informational level, the mathematical model for a double-layer feedforward ANN (with one hidden layer) can be written as:

$$\hat{y} = f_r^{<out>} \left( \sum_{j=1}^{N_{hidden}} W_{j,r}^{<out>} f_j \left( \sum_{i=1}^{N_{input}} W_{i,j} x_i + b_j \right) + b_r^{<out>} \right), \tag{9}$$

where x, $\hat{y}$ are vectors consisting of the input and output information of the ANNM; $j$ is a serial number of the neuron of the hidden layer ($N_{hidden}$ is a total number of neurons in the hidden layer); $i$ is a serial number of the ANNM input ($N_{input}$ is a total number of inputs); $r$ is a serial number of the neuron of the output layer; W, $W_{j,r}^{<out>}$ are the arrays consisting of synaptic weights of neurons of the hidden and output layers; $b$, $b_r^{<out>}$ are the vectors consisting of biases of synapses of neurons of the hidden and output layers; $f_r^{<out>}()$, $f()$ are the transfer functions of each neuron of the hidden and output layers.

The number of MLP inputs is 16 corresponding to four channels of the microfluidic chip, in each of them the times of spikes are registered during 50 ms (about 3-4 spikes). The number of neurons in a hidden layer is 8 corresponding to the maximum number of neurons that can be implemented using a single 16 × 16 memristor array. The number of output neurons is 4 corresponding to four sites of stimulation S1, ..., S4 in chamber A. Each neuron has saturating linear transfer functions, which are closer to real transfer functions implemented using differential amplifiers.

The performance of the ANNM during the training process is evaluated using the Mean Squared Error:

$$MSE = \frac{1}{H} \sum_{h=1}^{H} (y_h - \hat{y}_h)^2, \tag{10}$$

where $H$ is the total number of elements of the training, validation or test set, $y$ consists of target values.

The performance of the ANNM during the operation is evaluated using the Probability of Error, which indicates the number of correctly classified signal patterns compared to the total number of patterns:

$$P_{err} = \frac{E}{H} \cdot 100\%,\qquad\qquad(11)$$

where $E$ is the number of incorrectly recognized signal patterns.

The synthesis of tolerances for informational parameters and analysis of tolerances for physical parameters of the ANNM was performed at the permissible value of the probability of error ($X_p$) $P_{err} = 5\%$.

## 3    Results and discussion

### 3.1    ANNM training results

During the MLP training process, it is necessary to adjust synaptic weights to minimize the loss function to a given target value over a given number of epochs. However, during the training process, the comparable values of ANNM operation accuracy can be achieved with different sets of weights (Danilin et al., 2019). It depends on many reasons. For example, different weights for the same algorithm can be obtained by reaching different local minima starting from different initial values or set different hyperparameters before training.

This phenomenon is often insignificant for the digital implementation of ANN, in which the multiplication is only an arithmetic operation on binary numbers. However, for the ANNM hardware analog implementation, the multiplication is performed by the Ohm's law and variation of memristors resistances in comparison with nominal values cause errors of synaptic weights, which in turn can lead to significant computational errors.

For these reasons, the MLP was trained using a modified algorithm based on the truncated Newton method (Robinson, 2018). The modifications touched two steps of this method:

- When updating a weight at each epoch, the closest discrete value is assigned calculated according to discrete stable states of synapse memristors.
- If it is experimentally determined in advance that the memristor or both synapse memristors cannot change their state (stuck-at memristors), then the weight can either take a certain value from a limited set of discrete values or be constant during the training.

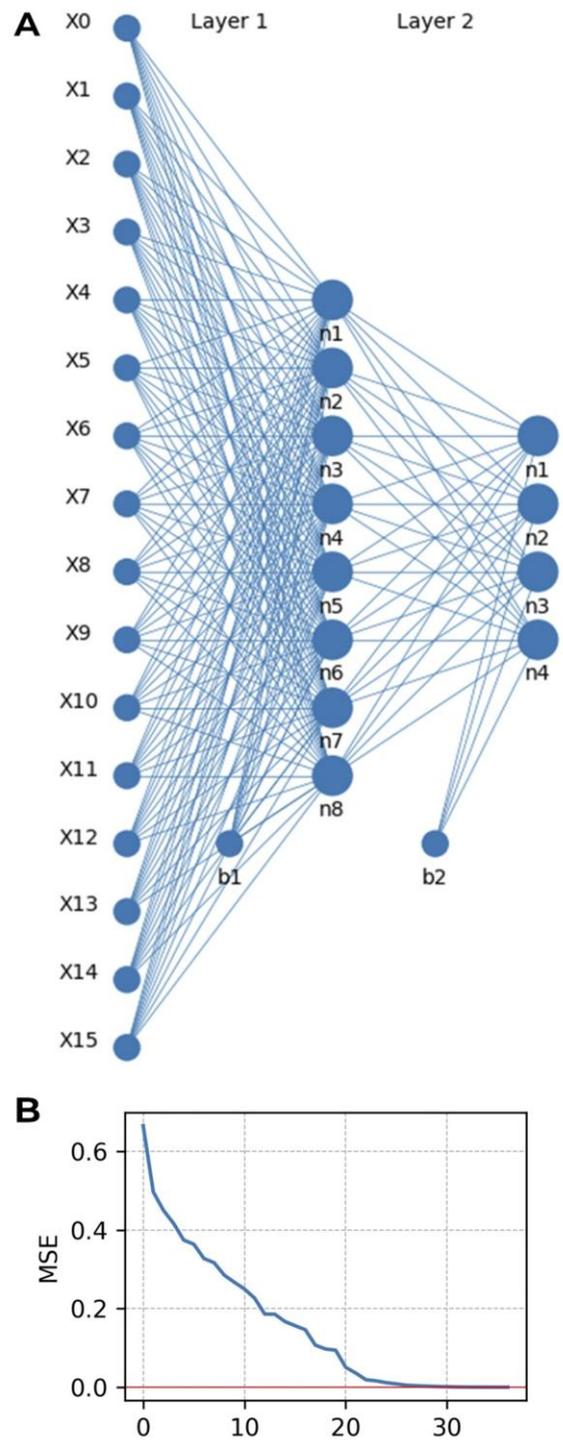

**Figure 7 |** Subsystem level modeling. **(A)** The structure of the double-layer feedforward ANNM. It has 16 inputs, 8 neurons in the hidden layer and 4 neurons in the output layer. Total number of synapses is 160 and memristors is 320. Biases are implemented in digital hardware for providing additional ability to tweak. **(B)** The learning curve for the MLP, which illustrates the training process with the use of the proposed algorithm.

A similar approach is described, for example, in papers (Yang et al., 2016; Bayat et al., 2018; Li et al., 2018) and makes it possible to provide the required accuracy of ANNM training.

The target value of MSE on a training dataset equals to 0.0001, the maximum number of training epochs is 10000. The probability of error calculated after training on the test set is less than 2%. The training curve is shown in **Figure 7B**.

Then, the values of synaptic weights obtained at the informational level should be presented in the form of electrical parameters of electronic elements of synapses at the device (physical) level. On the next step, the simulation of ANNM operation at the given values of limits of variations in values of the parameters of circuit elements was conducted. Detailed description of these investigations is given in **Supplementary Materials**.

### 3.2 ANNM simulation results

The limits of errors of MLP synaptic weights were determined during a simulation (see **Figure 8A**). Nominal values for circuit elements are $R_F = 100$ k$\Omega \pm 1\%$, $R_{M1} = 322.9$ k$\Omega \pm 20\%$, $R_{M2} = 12.2$ k$\Omega \pm 20\%$. Boundaries of errors ranges of weights were defined for 0.05-th and 99.5-th percentiles. This simulation was performed with the use of the model at the device level. Then the largest deviations for each weight were taken and set to the model at the subsystem level (see **Figure 8B**). Next, we performed a simulation of the ANNM in the presence of errors in the synaptic weights caused by the errors in circuit elements (see **Figure 8C**). After 10,000 repetitions of the experiment, the probability of error $P_{err}$ for the ANNM does not exceed 5%. This value was obtained on a test sample. For a more detailed analysis, additional modeling was carried out as described in detail in **Supplementary Materials**.

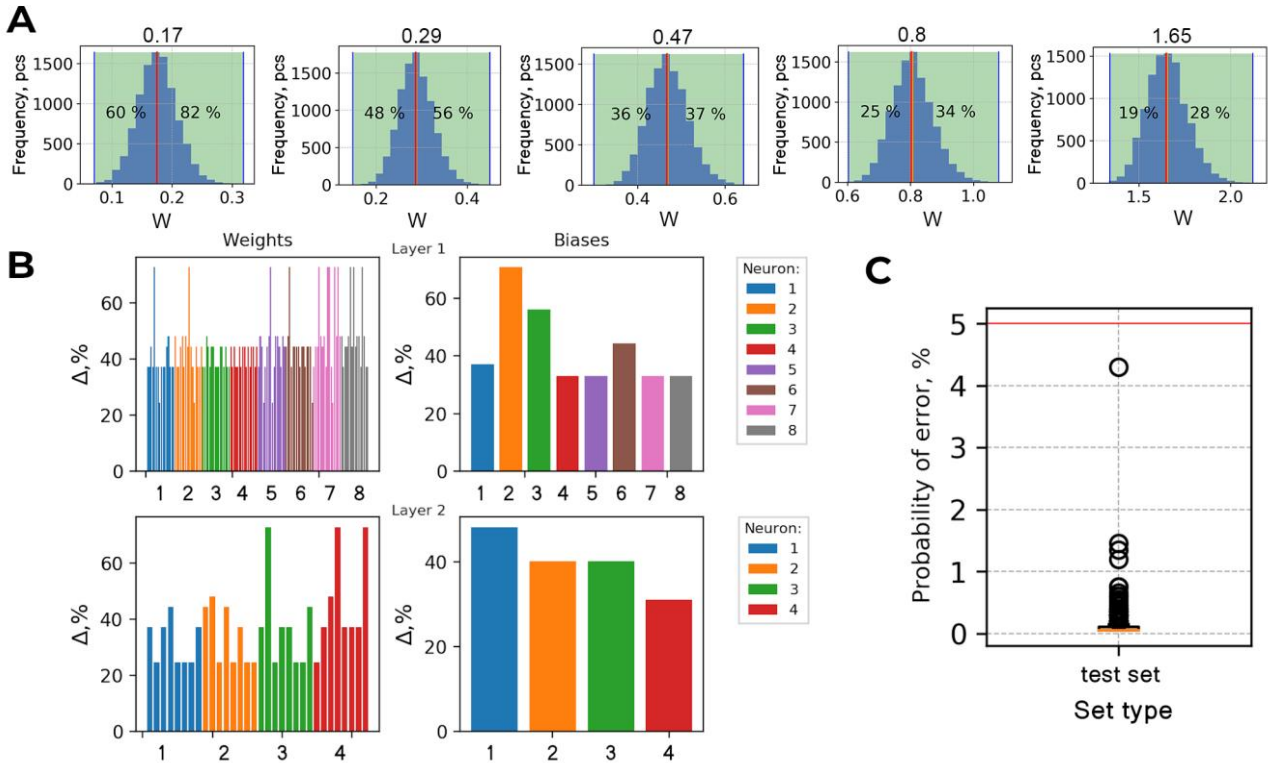

**Figure 8** | Simulation results. **(A)** PDFs for each weight ($W = 0.17, 0.29, 0.47, 0.80, 1.65$) after simulation of errors in circuit elements (memristors $R_{M1}$ and $R_{M2}$ and resistance $R_f$) of synapses. Red line is nominal value, $\Omega$. Blue lines are 0.05-th and 99.5-th percentiles. Green area is percentage ranges of errors in weights. **(B)** Limits of variations in weights for all ANNM components (for each weight (or bias) for each neuron (marked with different colors) in each layer). **(C)** The boxplot

shows the probability of ANNM error in the presence of errors in synaptic weights estimated on the test set. The simulation loop was carried out 10 000 times. It can be seen that these values do not exceed 5%.

The obtained probability of error is currently the best result achieved. Although the variation of parameters of memristive devices is a significant problem for constructing traditional neural networks that require an exact programmed value of synaptic weight, the stochastic nature of resistive switching will be useful in new generations of spiking neural networks with self-organization of weights based on the rich dynamics of memristor internal state.

As a result of the ANNM design, a set of models for different levels of the structural and functional hierarchy is obtained, which makes it possible to determine the nominal values of electronic elements of the neural network circuit as a component of the neural interface. As a result of simulation, the permissible limits of deviations of circuit elements parameters from the nominal values are assigned. As the next step, the prototyping and testing of the developed ANNM are planned.

Thus, the memristive devices of the $Au/Ta/ZrO_2(Y)/Ta_2O_5/TiN/Ti$ type are suitable in their electrophysical parameters and characteristics for the hardware implementation of ANNM, which is a component of a bidirectional adaptive neurointerface.

## 4    Conclusion

- The ANNM theory component is developed in part of determining and providing the required accuracy of synapses operation at the implementation of models in hardware. It includes a general approach, method and algorithm, which allow us to determine the permissible limits of the ANNM synapse errors. All developed models and algorithms are implemented in the form of Python-based software.
- The ANNM for a bidirectional adaptive neural interface with a signal recognition accuracy of at least 95% is designed. The ANNM project includes a set of structural-functional models necessary and sufficient to solve the tasks assigned in terms of reference.
- Applicability of passive memristive crossbars for the hardware implementation and real practical use of neuromorphic systems is demonstrated in the paper among the first along with Strukov and coauthors (Bayat et al., 2018), who implemented a multilayer perceptron on two cross- bar arrays of about the same size ($20 \times 20$). Cross-bars with a larger number of elements (Kataeva et al., 2019) are currently under development and CMOS integration, and the proposed general approach can be used for their implementation in neural networks.
- The article proposes an original circuit of a multilayer perceptron based on memristors. Its main advantages include easy scalability and the ability to adjust the range of weights by changing the values of resistors. The proposed circuit diagram of a neuron is distinguished by a circuit for switching bias voltages to provide a neuron programming mode in terms of neutralizing bypass currents. The difference from other solutions known from publications is the absence of switching elements in the circuits, in which the memristors are included. This minimizes the interference associated with operation of switching elements.
- The authors continue to work on the issue considered in the article. The next stage of the work is the hardware implementation of the developed and investigated models of ANNM, their prototyping and testing.

## 5    Author Contributions

All authors gave substantial contribution to the development of this work equally, drafting and revising it critically; furthermore, all Authors approved its final version for publication.

## 6    Funding

The research was supported by the Russian Science Foundation (grant No. 16-19-00144).

## 7    Conflicts of Interest

The authors declare that the research was conducted in the absence of any commercial or financial relationships that could be construed as a potential conflict of interest.

## 8    References

Bayat, F. M., Prezioso, M., Chakrabarti, B., Nili, H., Kataeva, I., and Strukov, D. (2018). Implementation of multilayer perceptron network with highly uniform passive memristive crossbar circuits. *Nat. Commun.* 9. doi:10.1038/s41467-018-04482-4.

Bernier, J. L., Ortega, J., Rojas, I., and Prieto, A. (2000a). Improving the tolerance of multilayer perceptrons by minimizing the statistical sensitivity to weight deviations. *Neurocomputing* 31, 87–103. doi:10.1016/S0925-2312(99)00150-2.

Bernier, J. L., Ortega, J., Rojas, I., Ros, E., and Prieto, A. (2000b). Obtaining fault tolerant multilayer perceptrons using an explicit regularization. *Neural Process. Lett.* 12, 107–113. doi:10.1023/A:1009698206772.

Boi, F., Moraitis, T., Feo, V. De, Diotalevi, F., Bartolozzi, C., Indiveri, G., et al. (2016). A bidirectional brain-machine interface featuring a neuromorphic hardware decoder. *Front. Neurosci.* 10. doi:10.3389/fnins.2016.00563.

Brivio, S., Frascaroli, J., Covi, E., and Spiga, S. (2019). Stimulated Ionic Telegraph Noise in Filamentary Memristive Devices. *Sci. Rep.* 9. doi:10.1038/s41598-019-41497-3.

Buccelli, S., Bornat, Y., Colombi, I., Ambroise, M., Martines, L., Pasquale, V., et al. (2019). A Neuromorphic Prosthesis to Restore Communication in Neuronal Networks. *iScience* 19, 402–414. doi:10.1016/j.isci.2019.07.046.

Chiolerio, A., Chiappalone, M., Ariano, P., and Bocchini, S. (2017). Coupling resistive switching devices with neurons: State of the art and perspectives. *Front. Neurosci.* 11. doi:10.3389/fnins.2017.00070.

Chua, L. (2018). Five non-volatile memristor enigmas solved. *Appl. Phys. A Mater. Sci. Process.* 124. doi:10.1007/s00339-018-1971-0.

Chua, L. O. (1971). Memristor—The Missing Circuit Element. *IEEE Trans. Circuit Theory* 18, 507–519. doi:10.1109/TCT.1971.1083337.

Danilin, S. N., Shchanikov, S. A., Bordanov, I. A., and Zuev, A. D. (2019). The influence of algorithms for tuning the parameters of neuromorphic systems on their fault tolerance. *J. Phys. Conf. Ser.* 1333, 032077. doi:10.1088/1742-6596/1333/3/032077.

Danilin, S. N., Shchanikov, S. A., and Galushkin, A. I. (2015). The research of memristor-based neural network components operation accuracy in control and communication systems. in *2015 International Siberian Conference on Control and Communications, SIBCON 2015 - Proceedings* (Institute of Electrical and Electronics Engineers Inc.). doi:10.1109/SIBCON.2015.7147034.

Danilin, S. N., Shchanikov, S. A., and Panteleev, S. V. (2016). Determining Operation Tolerances of Memristor-Based Artificial Neural Networks. in *2016 International Conference on Engineering and Telecommunication (EnT)* (IEEE), 34–38. doi:10.1109/EnT.2016.016.

Demin, V. A., Erokhin, V. V., Emelyanov, A. V., Battistoni, S., Baldi, G., Iannotta, S., et al. (2015). Hardware elementary perceptron based on polyaniline memristive devices. *Org. Electron.* 25, 16–20. doi:10.1016/j.orgel.2015.06.015.

Emelyanov, A. V., Lapkin, D. A., Demin, V. A., Erokhin, V. V., Battistoni, S., Baldi, G., et al. (2016). First steps towards the realization of a double layer perceptron based on organic memristive devices. *AIP Adv.* 6. doi:10.1063/1.4966257.

Emelyanov, A. V., Nikiruy, K. E., Demin, V. A., Rylkov, V. V., Belov, A. I., Korolev, D. S., et al. (2019). Yttria-stabilized zirconia cross-point memristive devices for neuromorphic applications. *Microelectron. Eng.* 215. doi:10.1016/j.mee.2019.110988.

Galushkin, A. I. (2007). *Neural network theory*. Springer Berlin Heidelberg doi:10.1007/978-3-540-48125-6.

Gryaznov, Y. G., Antonov, I., Kotina, A., Kotomina, V., Mikhailov, A., Sharapov, A., et al. (2018). Russian sertificate of state registration of integrated circuit topology No. 018630129. Topology of test chip with array of memristive microdevices No. 2018630123.

Gupta, I., Serb, A., Khiat, A., Zeitler, R., Vassanelli, S., and Prodromakis, T. (2016). Real-time encoding and compression of neuronal spikes by metal-oxide memristors. *Nat. Commun.* 7. doi:10.1038/ncomms12805.

Gupta, I., Serb, A., Khiat, A., Zeitler, R., Vassanelli, S., and Prodromakis, T. (2018). Sub 100 nW volatile nano-metal-oxide memristor as synaptic-like encoder of neuronal spikes. *IEEE Trans. Biomed. Circuits Syst.* 12, 351–359. doi:10.1109/TBCAS.2018.2797939.

Hamdioui, S., Aziza, H., and Sirakoulis, G. C. (2014). Memristor based memories: Technology, design and test. in *Proceedings - 2014 9th IEEE International Conference on Design and Technology of Integrated Systems in Nanoscale Era, DTIS 2014* (IEEE Computer Society). doi:10.1109/DTIS.2014.6850647.

Hogri, R., Bamford, S. A., Taub, A. H., Magal, A., Giudice, P. Del, and Mintz, M. (2015). A neuro-inspired model-based closed-loop neuroprosthesis for the substitution of a cerebellar learning function in anesthetized rats. *Sci. Rep.* 5. doi:10.1038/srep08451.

Kataeva, I., Ohtsuka, S., Nili, H., Kim, H., Isobe, Y., Yako, K., et al. (2019). Towards the development of analog neuromorphic chip prototype with 2.4M integrated memristors. in *Proceedings - IEEE International Symposium on Circuits and Systems* (Institute of Electrical and Electronics Engineers Inc.). doi:10.1109/ISCAS.2019.8702125.

Lanza, M., Wong, H. S. P., Pop, E., Ielmini, D., Strukov, D., Regan, B. C., et al. (2019). Recommended Methods to Study Resistive Switching Devices. *Adv. Electron. Mater.* 5. doi:10.1002/aelm.201800143.

Li, C., Belkin, D., Li, Y., Yan, P., Hu, M., Ge, N., et al. (2018). Efficient and self-adaptive in-situ learning in multilayer memristor neural networks. *Nat. Commun.* 9. doi:10.1038/s41467-018-04484-2.

Mehonic, A., Joksas, D., Ng, W. H., Buckwell, M., and Kenyon, A. J. (2019). Simulation of inference accuracy using realistic rram devices. *Front. Neurosci.* 13. doi:10.3389/fnins.2019.00593.

Mikhaylov, A. N., Morozov, O. A., Ovchinnikov, P. E., Antonov, I. N., Belov, A. I., Korolev, D. S., et al. (2018). One-Board Design and Simulation of Double-Layer Perceptron Based on Metal-Oxide Memristive Nanostructures. *IEEE Trans. Emerg. Top. Comput. Intell.* 2, 371–379. doi:10.1109/tetci.2018.2829922.

Park, S., Chu, M., Kim, J., Noh, J., Jeon, M., Hun Lee, B., et al. (2015). Electronic system with memristive synapses for pattern recognition. *Sci. Rep.* 5. doi:10.1038/srep10123.

Pershin, Y. V., and Di Ventra, M. (2020). On the validity of memristor modeling in the neural network literature. *Neural Networks* 121, 52–56. doi:10.1016/j.neunet.2019.08.026.

Pimashkin, A., Gladkov, A., Mukhina, I., and Kazantsev, V. (2013). Adaptive enhancement of learning protocol in hippocampal cultured networks grown on multielectrode arrays. *Front. Neural Circuits*. doi:10.3389/fncir.2013.00087.

Pimashkin, A., Kastalskiy, I., Simonov, A., Koryagina, E., Mukhina, I., and Kazantsev, V. (2011). Spiking signatures of spontaneous activity bursts in hippocampal cultures. *Front. Comput. Neurosci.* 5. doi:10.3389/fncom.2011.00046.

Robinson, S. M. (2018). *Numerical Optimization 2nd Edition*. doi:10.1007/978-0-387-40065-5.

Romano, D., Donati, E., Benelli, G., and Stefanini, C. (2019). A review on animal–robot interaction: from bio-hybrid organisms to mixed societies. *Biol. Cybern.* 113, 201–225. doi:10.1007/s00422-018-0787-5.

Schuman, C. D., Potok, T. E., Patton, R. M., Birdwell, J. D., Dean, M. E., Rose, G. S., et al. (2017). A Survey of Neuromorphic Computing and Neural Networks in Hardware. Available at: http://arxiv.org/abs/1705.06963 [Accessed July 13, 2019].

Serb, A., Corna, A., George, R., Khiat, A., Rocchi, F., Reato, M., et al. (2017). A geographically distributed bio-hybrid neural network with memristive plasticity. Available at: http://arxiv.org/abs/1709.04179 [Accessed August 25, 2019].

Shannon, R. (1975). *Systems Simulation: The Art and Science*. Prentice Hall.

Strukov, D. B., Snider, G. S., Stewart, D. R., and Williams, R. S. (2008). The missing memristor found. *Nature* 453, 80–83. doi:10.1038/nature06932.

Tihov, S., Belov, A., Korolev, D., Antonov, A., Sushkov, D., Pavlov, D., et al. (2020). Electrophysical characteristics of multilayer memristive nanostructures based on yttrium stabilized zirconium dioxide and tantalum oxide. *Tech. Phys.* 90, 298–304.

Torres-Huitzil, C., and Girau, B. (2017). Fault and Error Tolerance in Neural Networks: A Review. *IEEE Access* 5, 17322–17341. doi:10.1109/ACCESS.2017.2742698.

Vassanelli, S., and Mahmud, M. (2016). Trends and challenges in neuroengineering: Toward "intelligent" neuroprostheses through brain-"brain inspired systems" communication. *Front. Neurosci.* 10. doi:10.3389/fnins.2016.00438.

Xia, Q., and Yang, J. J. (2019). Memristive crossbar arrays for brain-inspired computing. *Nat. Mater.* 18, 309–323. doi:10.1038/s41563-019-0291-x.

Yang, C., Kim, H., Adhikari, S., and Chua, L. (2016). A Circuit-Based Neural Network with Hybrid Learning of Backpropagation and Random Weight Change Algorithms. *Sensors* 17, 16. doi:10.3390/s17010016.

Yao, P., Wu, H., Gao, B., Eryilmaz, S. B., Huang, X., Zhang, W., et al. (2017). Face classification using electronic synapses. *Nat. Commun.* 8. doi:10.1038/ncomms15199.

Yeung, D., Cloete, I., Shi, D., and Ng, W. (2010). *Sensitivity Analysis for Neural Networks*. Springer Berlin.

Zidan, M. A., Strachan, J. P., and Lu, W. D. (2018). The future of electronics based on memristive systems. *Nat. Electron.* 1, 22–29. doi:10.1038/s41928-017-0006-8.

# *Supplementary Material*

## 1    Synapse simulation results

Calculation of maximum synaptic weights $W$ for the developed version of ANNM hardware implementation is performed according to expression (S1). For example, at $R_F = 100$ kΩ, $R_{MAX}$ (HRS) = 300 kΩ, $R_{MIN}$ (LRS) = 10 kΩ, the range of synaptic weights is $W \in [0; 9.67]$. The maximum value of the memristor resistance $R_{M2}$ does not significantly affect the range of weights $W$. So, for example, when $R_{M2} = 100$ kΩ, $W_{MAX} = 9.00$, $R_{M2} = 200$ kΩ, $W_{MAX} = 9.50$, $R_{M2} = 300$ kΩ, $W_{MAX} = 9.67$. As can be seen in **Figure S1 A**, the change in $R_{M2}$ in15 times causes the change in $W_{MAX}$ only in 1.92 times.

$$W_{MAX} = R_F \cdot \frac{(R_{MAX} - R_{MIN})}{(R_{MAX} \cdot R_{MIN})} \tag{S1}$$

The minimum value of the memristor resistance $R_{M1}$ strongly affects the range of weights $W$. So, for example, when $R_{M1} = 20$ kΩ, $W_{MAX} = 4.67$, $R_{M1} = 10$ kΩ $W_{MAX} = 9.67$, $R_{M1} = 5$ kΩ $W_{MAX} = 19.67$. As can be seen in **Figure S1 B**, the change in $R_{M1}$ 29 times causes the change in $W_{MAX}$ in 841 times.

The resistance $R_F$ linear affects the weight coefficient $W$. $R_F$ can be used for tuning the ANNM, to be exact for adjusting maximum synaptic weight $W_{MAX}$ to a required value, under specific values of HRS and LRS of memristors that are used.

When changing the value of $R_{M1}$ with a fixed step $\Delta R_{M1,}$ the change in weight $W$ is nonlinear. For example, **Figure S1 C** shows that, when the resistance $R_{M1}$ is changed with increments of $\Delta R_{M1} = 5$ kΩ at $R_{M2} = 60$ kΩ, the synaptic weight changes nonlinearly and has 11 discrete values. This means that, if during training change the resistance of memristor with step $\Delta R_{M1} = 5$ kΩ, 5 values can be set for the range of weights $W \in [0; 1)$, 2 values can be set for the range $W \in [1; 2)$, and 4 values can be set for the range $W \in [2; 9)$. In **Figure S1 D**, the inverse relationship is plotted. It can be seen that the range of weights $W \in [0; 1]$ corresponds to the range of resistances $R_{M1} \in [60; 37.5]$ kΩ ($\Delta R_{M1} = 22.5$ kΩ), which is more than 37 % of the entire resistance range of the memristor. For $W = 7$, $R_{M1} = 11.5$ kΩ, and for $W = 8$, $R_{M1} = 10.3$ kΩ, i.e., a change in the weight $W$ by 1 corresponds to a change in resistance by 1.2 kΩ. This means that the larger the value of weight $W$, the smaller is the range of resistances available for programming.

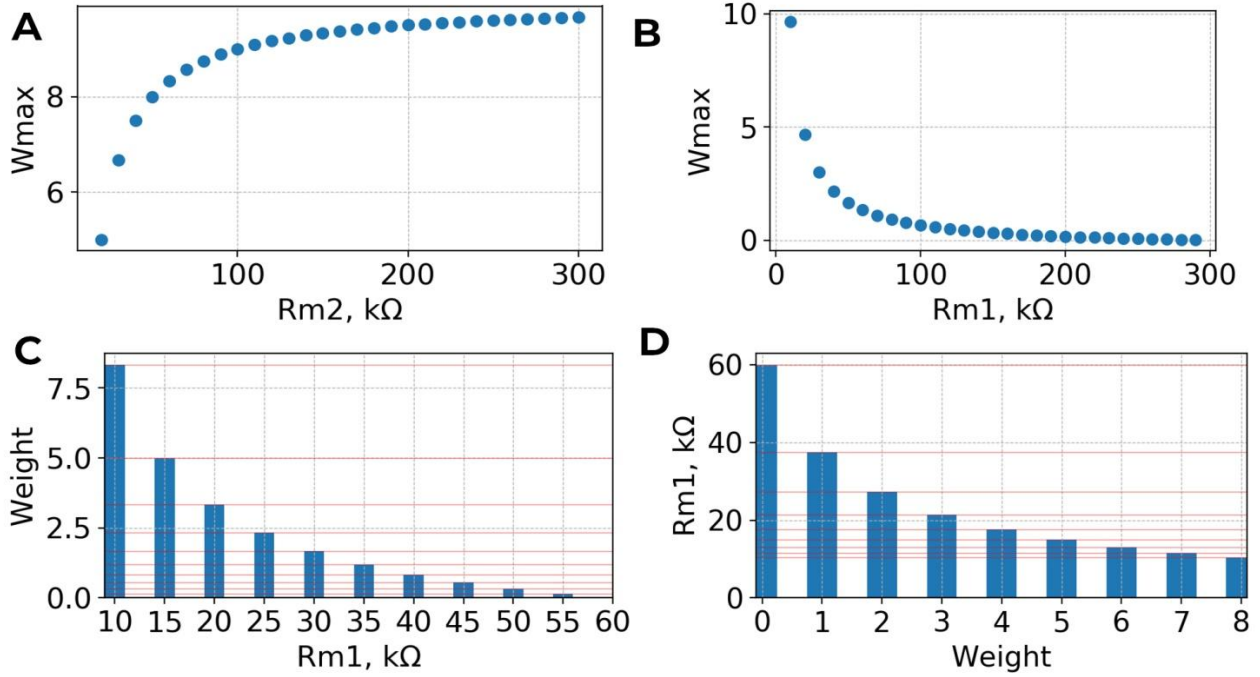

**Figure S1.** (**A**) Maximum weight $W_{MAX}$ versus maximum resistance of memristor $R_{M2}$. It shows that increasing the HRS of the synapse memristor from 100 to 300 kΩ, the maximum weight value does not change significantly, only on 0.67 (from 9.0 to 9.67). (**B**) Maximum weight $W_{MAX}$ versus maximum resistance of memristor $R_{M2}$. It shows that by decreasing the LRS of the synapse memristor from 20 to 5 kΩ, the weight changes 4 times from 4.67 to 19.67. (**C**) and (**D**) show the direct and inverse dependence of the weight on the resistance of memristor $R_{M1}$, respectively. As can be seen in (**C**) if one program memristors with a discrete step, then for smaller values of the weights there will be more discrete values, and for larger values of the weights there will be less discrete values. On the one hand, this is a positive side, because during the training process most algorithms strive to select lower values of weights, therefore, the ANNM inherent accuracy with lower values of weights will be better. On the other hand, as it is shown further in **Figure 10**, memristors non-idealities have a greater effect on lower weights than higher ones. As can be seen in (**D**) if it is necessary to program weights with a linear step, then the resistance of memristors will have to be changed nonlinearly, and for smaller values of resistance the difference will be less than for larger ones.

As a result of the presence of variations in circuit elements of a ANNM synapse, variations in weights $W$ occur (see **Figure S2 A**). The limits of these variations are calculated using simulation. For this purpose, it is necessary to set for all circuit elements of synapses the ranges of deviations according to the normal law, perform simulation and evaluate their effect on synaptic weights $W$.

In **Figure S2 B** it can be seen that, under the normal law of distribution of the nominal resistance values ($R_F = 100$ kΩ, $R_{M1} = 10$ kΩ, $R_{M2} = 300$ kΩ) of circuit elements at $\delta R_F = \pm 1\%$, $\delta R_{M1} = \delta R_{M2} = \pm 10\%$ in the worst case $\delta W = \pm 95\%$, and at $\delta R_{M1} = \delta R_{M2} = \pm 20$ % in the worst case $\delta W = \pm 160\%$. In **Figure S2 C** it can be seen that, with a larger value of $R_{MAX} = R_{M2} = 300$ kΩ, the maximum error value is lower (from 18% at $\delta R_M = 10\%$ to 47% at $\delta R_M = 20\%$).

In **Figure S2 D** it can be seen that the ANNM with nominal values of the parameters of circuit elements works with permissible accuracy starting from 7 discrete states of memristors.

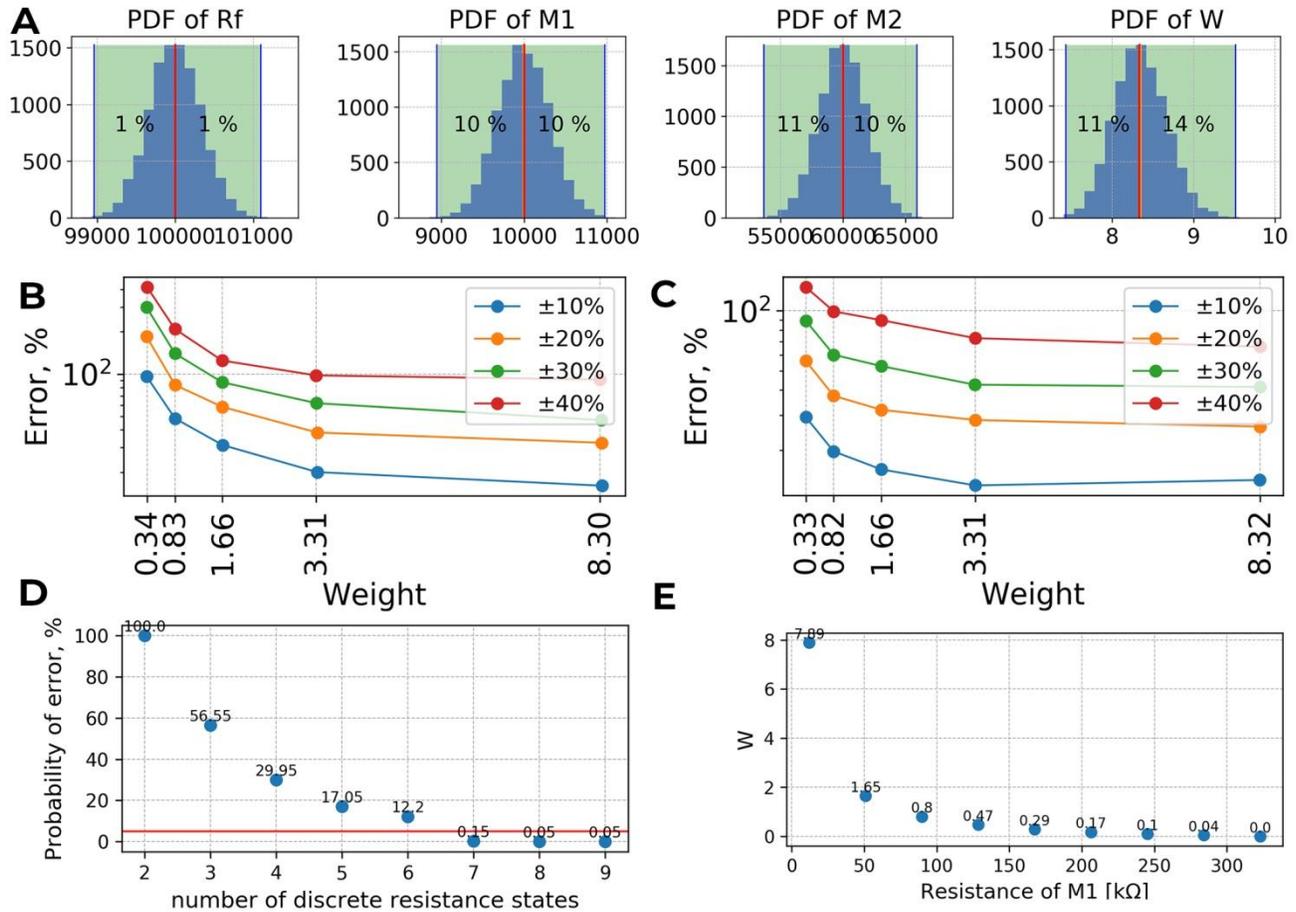

**Figure S2.** The results of simulation of ANNM synapses operation. (**A**) An example of the influence of variations in circuit elements ($R_F$, $R_{M1}$, $R_{M2}$) on the synaptic weight $W$. The limits of these variations are highlighted in green: the left border is 0.05-th percentile, the right border is 99.95-th percentile. The orange line is a mean value. The red line is the nominal value of a weight $W$. (**B**), (**C**) show the limits of errors of weights caused by different ranges of errors of memristors $\delta M_1 = \delta M_2 =$ [10, 20, 30, 40] % and for different values of $R_{MAX}$ (left panel: $R_{MAX} = 100$ kΩ, right panel: $R_{MAX} =$ 300 kΩ). (**D**) The probability of ANNM error when rounding weights to discrete states of memristor resistances. (**E**) Weights for 9 discrete resistance states of memristors.

## 2    Detailed analysis of ANNM simulation results

For more detailed analysis, the test set was divided into 2 parts − the first part includes only response patterns after stimuli from sites S1, ..., S4, and the second − only Sr signal patterns. As can be seen in **Figure S3 A,B** for both types of patterns, the probability of error does not exceed 5%. As the analysis of the ANNM output signal shows, the greatest errors are observed for neuron 3 (see **Figure S3 C**), when the signals of pattern S4 are at the input, and for neuron 2, when extraneous signals Sr are at the input (see **Figure S3 D**).

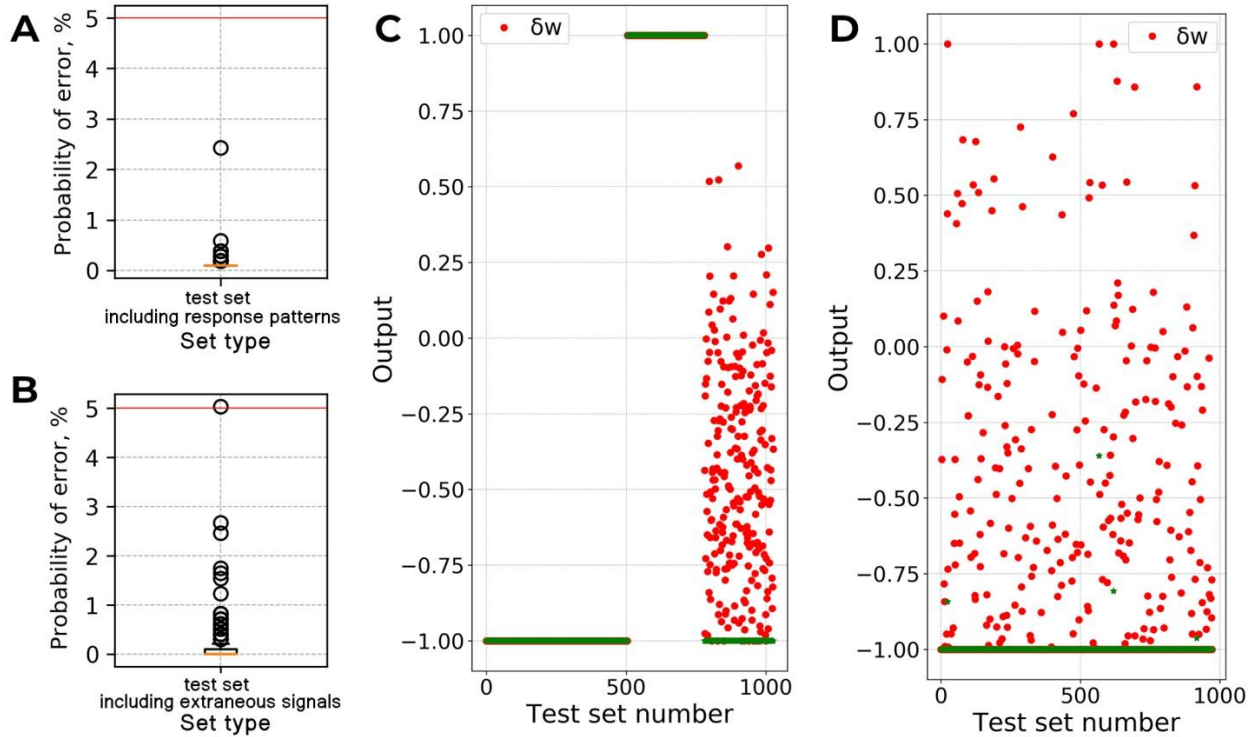

**Figure S3.** Results of the detailed analysis of the ANNM operation. Boxplots (**A**) and (**B**) illustrate probabilities of error of the ANNM in the presence of variations in synaptic weights calculated on the test set for responses after stimuli applied to sites S1, ..., S4 and for extraneous signals Sr, respectively. It can be seen that these values does not exceed 5% both for signals after target stimuli and for extraneous signals. In (**C**) (for the third neuron) and (**D**) (for the second neuron), the target output signals from the test set are shown in green, and ANNM output signals obtained during the simulation are shown in red.